# Electrical Switching of the Edge Current Chirality in Quantum Anomalous Hall Insulators

Wei Yuan[1,3], Ling-Jie Zhou[1,3], Kaijie Yang[1,3], Yi-Fan Zhao[1], Ruoxi Zhang[1], Zijie Yan[1], Deyi Zhuo[1], Ruobing Mei[1], Moses H. W. Chan[1], Morteza Kayyalha[2], Chao-Xing Liu[1], and Cui-Zu Chang[1]

[1]Department of Physics, The Pennsylvania State University, University Park, PA 16802, USA

[2]Department of Electrical Engineering, The Pennsylvania State University, University Park, Pennsylvania 16802, USA

[3]These authors contributed equally: Wei Yuan, Ling-Jie Zhou, Kaijie Yang

Corresponding authors: cxl56@psu.edu (C.-X. L.); cxc955@psu.edu (C.-Z. C.)

**Abstract: A quantum anomalous Hall (QAH) insulator is a topological state of matter, in which the interior is insulating but electrical current flows along the edges of the sample, in either clockwise (right-handed) or counter-clockwise (left-handed) direction dictated by the spontaneous magnetization orientation [1-7]. Such chiral edge current (CEC) eliminates any backscattering, giving rise to quantized Hall resistance and zero longitudinal resistance. In this work, we fabricate mesoscopic QAH sandwich (i.e. magnetic topological insulator (TI)/TI/magnetic TI) Hall bar devices and succeed in switching the CEC chirality in QAH insulators through spin-orbit torque (SOT) by applying a current pulse and suitably controlled gate voltage. The well-quantized QAH states with opposite CEC chiralities are demonstrated through four- and three-terminal measurements before and after SOT switching. Our theoretical calculations show that the SOT that enables the magnetization switching can be generated by both bulk and surface carriers in QAH insulators, in good agreement with experimental observations.**

**Current pulse-induced switching of the CEC chirality in QAH insulators will not only advance our knowledge in the interplay between magnetism and topological states but also expedite easy and instantaneous manipulation of the QAH state in proof-of-concept energy-efficient electronic and spintronic devices as well as quantum information applications.**

**Main text:** The success of the next-generation quantum material-based electronic and spintronic devices hinges on the creation of a reliable material platform with robust and low-energy consumption quantum states that are feasible for manipulations and measurements. The topological states of matter provide an appealing opportunity due to their intrinsic protection against impurity scattering [8,9]. The quantum anomalous Hall (QAH) state is one example of the topological states [1-7]. It possesses quantized Hall resistance with dissipationless chiral edge current (CEC), similar to that seen in the integer quantum Hall effect [10] but without the need for an external magnetic field. The QAH effect was first discovered in magnetically doped topological insulators (TI), specifically Cr- and V-doped $(Bi,Sb)_2Te_3$ thin films [6,7]. More recently, the zero magnetic field QAH effect has also been realized in mechanically exfoliated intrinsic magnetic TI $MnBi_2Te_4$ flakes with odd number layers [11], and manually assembled moiré materials formed from graphene [12] or transition metal dichalcogenides [13]. The resistance-free QAH-CEC has been predicted to have potential applications for next-generation of energy-efficient electronics and spintronics as well as quantum information applications [1].

In QAH insulators, the edge states are spin-polarized, and their chiralities are determined by internal magnetization, which leads to a unique feature of locking between the magnetization direction and the CEC propagating direction (i.e. chirality) (Figs. 1a and 1b).

The Hall resistance can be used as a probe to identify both magnetization direction and CEC chirality. To date, the CEC chirality (i.e. right- or left-handed propagating directions) in well-quantized QAH insulators can be controlled only by sweeping back and forth the external magnetic field $\mu_0 H$ (Refs. [1,6,7,14-18]). However, this is a cumbersome operation procedure in real CEC-based electronic devices. The ability to switch the CEC chirality in QAH insulators instantaneously without sweeping the external magnetic field is essential for the development of CEC-based computation and information technologies. In QAH insulators, the locking between magnetism and CEC chirality creates an opportunity to electrically switch the CEC chirality by reversing the magnetization direction through spin-orbit torque (SOT) [19,20]. The current-driven SOT has been proven to be an efficient mechanism for magnetization switching in TI/ferromagnet heterostructures [21-23], magnetic TI films [24], and magnetic TI/TI bilayer heterostructures [25,26]. However, SOT switching has not been demonstrated in any magnetic TI films/heterostructures that can host the QAH state. To realize electrical SOT switching in QAH insulators, the sample needs to satisfy two conditions: (*i*) the QAH structure can provide a large SOT effect, and (*ii*) the current density in QAH insulators can be high enough for magnetization switching. Condition (*i*) can be achieved in QAH sandwich heterostructures, in which an asymmetric potential between the top and bottom surfaces can be created by applying a gate voltage $V_g$ (Ref. [27]). Condition (*ii*) can be achieved by fabricating mesoscopic QAH sandwich Hall bar devices with micrometer size, which increases the current density. Since the top and bottom magnetic TI layers in QAH sandwiches have similar magnetic moments, these two magnetically doped TI layers switch their magnetization simultaneously. After SOT magnetization switching, the QAH insulator with opposite CEC chirality is induced.

In this work, we fabricate QAH sandwich (i.e. magnetic TI/TI/magnetic TI) Hall bar devices with the width of 1~10 μm using electron-beam lithography (Methods). By performing four- and three-terminal transport measurements [28], we demonstrate the electrical switching of the CEC chirality by employing SOT in QAH sandwiches. We find that the $C = 1$ QAH insulator with the right-handed CEC can be switched to the $C = -1$ QAH insulator with the left-handed CEC by applying a current pulse under an in-plane magnetic field. Since the SOT magnetization switching ratio is easily greater than the quantum percolation threshold of magnetic domains [29,30], a well-quantized QAH state with the opposite CEC chirality usually appears after the magnetization switching. This unique property makes the electrical switching of CEC chirality robust and reliable in real QAH electronic devices. We further theoretically study the SOT in QAH insulators, which can be induced by the inverse spin-galvanic effect from both bulk and surface carriers. Our calculations show that the first bulk valence quantum well sub-bands contribute the same sign to SOT as the hole bands of surface states, consistent with the experimental observations.

The QAH insulators used in this work are 3 quintuple layers (QL) Cr-doped $(Bi, Sb)_2Te_3$ /4 QL $(Bi, Sb)_2Te_3$ /3 QL Cr-doped $(Bi, Sb)_2Te_3$ sandwich heterostructures (referred to as 3-4-3 QAH sandwich), which are synthesized in a molecular beam epitaxy (MBE) chamber (Omicron Lab10) with the vacuum better than $2\times10^{-10}$ mbar. The electrical transport measurements are carried out in a Physical Property Measurement System (Quantum Design DynaCool, 1.7 K, 9 T) with a horizontal rotator and in a dilution refrigerator (Bluefors, 10 mK) cryostat with a vector magnet (9-1-1 T). Six-terminal Hall bars with bottom gate electrodes are used for electrical transport studies. More details about the MBE growth, device fabrication, and electrical transport measurements can be found in Methods.

We first assess the QAH state in a 3-4-3 sandwich Hall bar device with a width of ~5 μm (Device A, Extended Data Fig. 1) by performing electrical transport measurements at $T = 20$ mK. Figure 1c shows the magnetic field $\mu_0H$ dependence of the Hall resistance $\rho_{yx}$ and the longitudinal resistance $\rho_{xx}$ at the charge neutral point $V_g = V_g^0 = +0.8$ V. For positive magnetization $M > 0$, $\rho_{yx}$ has a quantized value of ~ +0.9919 $h/e^2$, and $\rho_{xx}$ ~ 0.0029 $h/e^2$ (~74 Ω) under zero magnetic field, corresponding to the $C = 1$ QAH state with right-handed CEC (Fig. 1a). By employing an external magnetic field to switch the magnetization polarity, i.e. for negative magnetization $M < 0$, $\rho_{yx}$ displays a negative quantized value of ~ -0.9954 $h/e^2$, and $\rho_{xx}$ ~ $7\times10^{-5}$ $h/e^2$ (~2 Ω) under zero magnetic field, giving rise to the $C = -1$ QAH state with left-handed CEC (Fig. 1b). The presence of the good QAH state is further validated by the gate voltage $V_g$ dependence of $\rho_{yx}$ and $\rho_{xx}$ at zero magnetic field [$\rho_{yx}(0)$ and $\rho_{xx}(0)$] (Fig. 1d). We set the initial state at $M > 0$, $\rho_{yx}(0)$ exhibits a distinct plateau at a positive quantized value of ~ 0.9978 $h/e^2$, concomitant with $\rho_{xx}(0)$ ~ 0.0028 $h/e^2$, centered at $V_g = V_g^0$ (Fig. 1d). Therefore, the initial state is the $C = 1$ QAH state with right-handed CEC (Fig. 1a). By employing the large SOT effect in topological materials [21-26], we will investigate if the QAH-CEC chirality can be electrically switched by applying a current pulse.

In a magnetic TI sandwich heterostructure near the insulating regime, an electrical current $J_e$ along the $x$-axis can generate spin accumulation $S_y$ along the $y$-axis at the magnetic TI/TI interface and exerts a damping-like SOT $\tau_{SO} \propto M \times (M \times S_y\hat{y})$ (Refs. [21-26]; Fig. 2a) on the magnetic moment $M$. Therefore, when an in-plane magnetic field $\mu_0H_{//}$ is applied parallel (antiparallel) to the electrical current direction, $\tau_{SO}$ can be described by an effective field $\mu_0H_{SO} \propto S_y\hat{y} \times M$, which drives $M$ to rotate from positive to negative (negative to positive) in a QAH sandwich heterostructure (Fig. 2a). Since the carriers on its two TI/TI interfaces have opposite spin polarization, $\tau_{SO}$ from two symmetry-equivalent surfaces of the QAH sandwich cancels each other (Fig. 2a). Therefore, to induce SOT in QAH insulators, we need to introduce an asymmetric potential to the two surfaces of the QAH sandwich samples via an external gate voltage $V_g$ (Fig. 2b).

The application of a single bottom gate $V_g$ inevitably induces an asymmetric potential in a QAH sandwich heterostructure [27]. Moreover, the introduction of some dissipative bulk conducting channels may also enhance the SOT effect through the spin Hall effect [20]. Therefore, we first tune $V_g$ to make $\rho_{yx}(0) \sim +0.27$ $h/e^2$ ($\sim +7.00$ k$\Omega$) when the chemical potential crosses both bulk valence bands and valence bands (Fig.2b). By injecting a series of current pulses, we examine the SOT magnetization switching at $\rho_{yx}(0) \sim +0.27$ $h/e^2$ under an in-plane magnetic field $\mu_0 H_{//} = \pm 0.05$ T (Fig. 2c). We note that $\mu_0 H_{//} = \pm 0.05$ T is chosen because it is much smaller than the anisotropy field $K \sim 0.7$ T of our QAH sandwiches (Extended Data Figs. 2 and 3). The values of $\rho_{yx}$ shown in Fig. 2c are measured at an excitation current of ~1 nA after injecting each current pulse. We find that the switching polarity is determined by the current pulse and the in-plane magnetic field directions, consistent with the SOT geometry (Fig. 2a). The current pulse $|I_{pulse}|$ ~200 μA is sufficient for the SOT magnetization switching in Device A ($w$ =5 μm). The corresponding current pulse density $J_{pulse}$ is shown on the upper horizontal axis of Fig. 2c. The critical current density for SOT magnetization switching is found to be ~ $4 \times 10^9$ A·m$^{-2}$, which is in the same order as that in topological materials but a few orders smaller than that in heavy metal/ferromagnet heterostructures [20,22,31,32].

Next, we employ the SOT effect induced by both bulk and surface carriers in magnetic TI sandwiches to realize the electrical switching of the CEC chirality. We start with the $C = 1$ QAH state with right-handed CEC as the initial state (Figs. 1a and 1c). As noted above, a large SOT can be induced by tuning $V_g$ to cross both bulk valence bands and surface states. We do the first switching at $(V_g - V_g^0) \sim -33$ V and $\rho_{yx}(0) \sim +0.2789$ $h/e^2$. By applying a current pulse $I_{pulse} \sim -200$ μA under $\mu_0 H_{//} = +0.05$ T, $\rho_{yx}(0)$ is switched to be ~ $-0.2321$ $h/e^2$, indicating the majority of magnetization is switched from positive to negative direction. Under zero magnetic field, we use $V_g$ to tune the chemical potential back to the magnetic exchange gap and find that $\rho_{yx}(0)$ displays a plateau with a negative quantized value of ~ -0.9836 $h/e^2$, concomitant with $\rho_{xx}(0) \sim 0.0229$ $h/e^2$ at $V_g = V_g^0$ (Fig. 2d). After the first switching, the sample behaves as a $C = -1$ QAH insulator with left-handed CEC (Fig. 1b), which is further validated by measuring the initial magnetization curves at $V_g = V_g^0$ (Fig. 2g). We next do the second and third switching at $|\rho_{yx}(0)| \sim 0.27\ h/e^2$ by applying a current pulse $I_{\text{pulse}} \sim +200\ \mu\text{A}$ and $I_{\text{pulse}} \sim -200\ \mu\text{A}$ under $\mu_0 H_{//} = +0.05$ T, respectively. We find that $\rho_{yx}(0) \sim +0.9833\ h/e^2$ and $\rho_{xx}(0) \sim 0.0163\ h/e^2$ after the second switch and $\rho_{yx}(0) \sim -0.9953\ h/e^2$ and $\rho_{xx}(0) \sim 0.0476\ h/e^2$ after the third switch at $V_g = V_g^0$, respectively (Figs. 2e and 2f). These observations confirm that the sample behaves as a $C = 1$ QAH insulator with right-handed CEC after the second switch and a $C = -1$ QAH insulator with left-handed CEC after the third switch, respectively. Both CEC chiralities are validated by measuring the initial magnetization curves at $V_g=V_g^0$ (Figs. 2h and 2i). For Device A, the SOT magnetization switching ratio at $|\rho_{yx}(0)| \sim 0.27\ h/e^2$ by applying $|I_{\text{pulse}}| \sim 200\ \mu\text{A}$ under $\mu_0 H_{//} = +0.05$ T is estimated to be ~83.8% (Extended Data Fig. 3), but the well-quantized QAH insulator emerges by tuning gates after each switch. This behavior is a result of the quantum percolation property in QAH insulators [29,30].

To further demonstrate the electrical switching of CEC chirality in QAH insulators, we perform three-terminal measurements on Device A after each switch (Fig. 3a). Figures 3b and 3c show the $\mu_0 H$ dependence of the three-terminal resistances $\rho_{14,13}$ and $\rho_{14,15}$ at $T = 20$ mK and $V_g = V_g^0$. At zero magnetic field, $\rho_{14,13} \sim h/e^2$ and $\rho_{14,15} \sim 0$ for $M > 0$, corresponding to $C = 1$ QAH insulator with right-handed CEC (Fig. 1a). For $M < 0$, $\rho_{14,13} \sim 0$ and $\rho_{14,15} \sim h/e^2$, corresponding to $C = -1$ QAH insulator with left-handed CEC (Fig. 1b). The $(V_g - V_g^0)$ dependence of $\rho_{14,13}$ and $\rho_{14,15}$ at zero magnetic field [$\rho_{14,13}(0)$ and $\rho_{14,15}(0)$] of the initial QAH state and the QAH states after the first and second switches are shown in Figs. 3d to 3f. At $V_g = V_g^0$, $\rho_{14,13}(0) \sim h/e^2$ and $\rho_{14,15}(0) \sim 0$ validate both the initial state and the state after the second switch are the $C = 1$ QAH insulator with right-handed CEC, $\rho_{14,13}(0) \sim 0$ and $\rho_{14,15}(0) \sim h/e^2$ validate the state after the first switch is the $C = -1$ QAH insulator with left-handed

CEC. Therefore, we establish that the QAH-CEC chirality can be switched by an excursion of a current pulse under an in-plane magnetic field through SOT induced by both bulk and surface carriers (Extended Data Fig. 4).

In addition to SOT generated by both bulk and surface carriers, we further employ the SOT effect induced by only surface carriers to switch the CEC chirality in QAH sandwich devices. We apply a current pulse on magnetic TI sandwich heterostructures at $V_g = V_g^0$ under an in-plane magnetic field (Fig. 4a) and investigate if the $C = 1$ QAH insulator can be switched to the $C = -1$ QAH insulator (Fig. 4b). For Device A, $\rho_{yx}(0) \sim +1.0005 h/e^2$ and $\rho_{xx}(0) \sim 0.0220 h/e^2$ at $V_g = V_g^0$, corresponding to the right-handed CEC. By applying $I_{pulse} \sim -200$ μA under $\mu_0 H_{//} = +0.05$ T at $V_g = V_g^0$, $\rho_{yx}(0)$ is found to be $\sim -0.4552 h/e^2$ and $\rho_{xx}(0)$ is $\sim 0.6166 h/e^2$. After the switch, we tune the sample back to the QAH regime with $\rho_{yx}(0) \sim -1.0008 h/e^2$ and $\rho_{xx}(0) \sim 0.0564 h/e^2$ at $V_g = V_g^0$, corresponding to the left-handed QAH-CEC (Fig. 4c). We replicate this phenomenon in a QAH sandwich Hall bar device with a width $w \sim 2$ μm (Device B, Extended Data Figs. 6 and 7). In Device B, we use $I_{pulse} \sim -100$ μA due to its narrower width (Extended Data Fig. 9). $\rho_{yx}(0)$ is $\sim +0.9999 h/e^2$ and $\sim -0.6005 h/e^2$ before and after the switching process, respectively. By tuning $V_g$, $\rho_{yx}(0)$ is $\sim -0.9865 h/e^2$ at $V_g = V_g^0$, concomitant with $\rho_{xx}(0) \sim 0.0072 h/e^2$, verifying the left-handed CEC chirality after switch (Fig. 4d). Therefore, we also realize the electrical switching of the QAH-CEC chirality through surface carriers induced SOT when the magnetic TI sandwich is near the QAH regime.

We next discuss two possible origins for the electrical switching of the CEC chirality at $V_g = V_g^0$. First, we find that the sample is usually tuned away from the QAH regime after the injection of the current pulse at $V_g = V_g^0$, which in turn induces an asymmetric potential on two surfaces of the QAH samples [27]. After SOT-induced switching, $\rho_{yx}(0)$ is $\sim -0.4552 h/e^2$ and $\sim -0.6005 h/e^2$ for Devices A and B, respectively (Figs. 4c and 4d). Both values indicate that the SOT induced by surface carriers plays a dominant role in magnetization and CEC chirality switching [1]. Second, the hundreds of μA current pulse is much larger than the breakdown current (tens to hundreds of nA) of the QAH insulators[26,33,34], so the QAH state must have been broken down after injecting the current pulse at $V_g = V_g^0$. A finite SOT is then likely created and induces magnetization switching. Moreover, we note that the quantum percolation property of the QAH insulators greatly facilitates the electrical switching of the CEC chirality (Extended Data Fig. 8). Provided that the SOT magnetization switching ratio through bulk and/or surface carriers generated SOT is greater than the quantum percolation threshold of magnetic domains in magnetic TI sandwiches [29,30], a well-quantized QAH state appears after each switch (Fig. 4b).

To support our experimental observations, we calculate the spin polarization responsible for SOT in QAH sandwiches and examine the roles of the bulk and surface carriers in inducing SOT. We model the QAH sandwiches of a thickness $L = (L_1 + L_2)$ (Fig.5a inset) with a four-band Hamiltonian[17,35]

$$H(\boldsymbol{k}_\parallel, -i\partial_z) = \left(\mathcal{D}_0(z) - D_1 \partial_z^2 + D_2 k_\parallel^2\right)\sigma_0 \tau_z + A_0\left(k_y \sigma_x - k_x \sigma_y\right)\tau_x - iB_0 \partial_z \sigma_0 \tau_y$$
$$+ \left(-C_1 \partial_z^2 + C_2 k_\parallel^2\right)\sigma_0 \tau_0 + \mathcal{M}(z)\sigma_z \tau_0 + V(z)\sigma_0 \tau_0,$$

where $\tau_{x,y,z}$ act on the orbital space and $\sigma_{x,y,z}$ act on the spin-1/2 space. The parameters are taken to be $D_1 = 6.86$ eVÅ$^2$, $D_2 = 44.5$ eVÅ$^2$, $A_0 = 3.33$ eVÅ, and $B_0 = 2.26$ eVÅ. $C_1 = -6$eVÅ$^2$ and $C_2 = 30$eVÅ$^2$ are used to tune the relative position of Dirac points and bulk valence bands. $\mathcal{M}(z)$ is the magnetization from Cr doping and $V(z) = V_0 \times (z/L - 1/2)$ is the asymmetric potential along the $z$-direction with $V_0 = 0.05$ eV. The QAH sandwich with magnetic dopants in the top and bottom surface layers is modeled by $\mathcal{D}_0(z) = \begin{cases} D_0 & L_1 \leq z \leq L_2 \\ D_0 + \Delta D_0 & 0 < z < L_1 \text{ or } L_2 < z < L \end{cases}$ and $\mathcal{M}(z) = \begin{cases} 0 & L_1 \leq z \leq L_2 \\ M & 0 < z < L_1 \text{ or } L_2 < z < L \end{cases}$ with $D_0 = -0.28$ eV and $\Delta D_0 = 0.26$ eV. For the 3-4-3 QAH sandwich heterostructure used in this work, $L_1 = 3$ nm and $L_2 = 7$ nm (Fig. 5a inset). The SOT $\tau_{so}$ is usually generated by spin polarization $S$. There are two origins of the spin polarization in a QAH sandwich [20], the traditional inverse spin-galvanic effect (ISGE) given by $S_{ISGE} \sim (\hat{z} \times E)$ (Refs. [21,22,24,26], Fig. 2a), and the surface Hall current (i.e. the magnetoelectric coupling) given by $S_{ME} \sim M_z \hat{z} \times (\hat{z} \times E)$ (Ref. [36,37], Fig. S4), both of which can be evaluated from the response function $\chi_{ij}$ that connects spin polarization to electric fields through $S_i = \sum_j \chi_{ij} E_j$ with $i, j = x, y, z$ (Methods). Magnetic switching can be caused by only the damping-like torque, either from $S_{ISGE}$ via $\tau_{so,ISGE} \propto M \times (M \times S_{ISGE})$ or $S_{ME}$ via $\tau_{so,ME} \propto M \times S_{ME}$. For the experimentally relevant ranges of the values of relaxation time, we find $S_{ISGE}$ is larger than $S_{ME}$, and thus we focus on $S_{ISGE}$ and $\tau_{so,ISGE}$ that come from $\chi_{yx}$. $\tau_{so,ME}$ and $S_{ME}$ from $\chi_{xx}$ are also discussed in Supplementary Information, which can also contribute to SOT magnetization switching.

Figures 5a and 5b show the energy spectrum and the response function $\chi_{yx}$ as a function of magnetization $M$ and the chemical potential $E_F$ for a fixed asymmetric potential in the 3-4-3 QAH sandwich, respectively. For $E_F > -0.1$ eV, the carriers near $E_F$ primarily come from the two surface states. A nonzero $\chi_{yx}$ is found and its sign is reversed from electron- to hole-doped regimes. This sign change can be understood from a two-surface-state model (Methods and Fig. 5c). For a specific Dirac surface state, $\chi_{yx}$ comes from the helical spin textures of electrons at the chemical potential and vanishes when $E_F$ is within the magnetic exchange gap. Because of opposite helical spin textures, $\chi_{yx}$ from the top and bottom surface states have opposite signs and thus tend to cancel each other [21-26] (Fig. 5c). When an asymmetric potential $V(z)$ is introduced, the relative shift between the top and bottom surface states leads to a net contribution of $\chi_{yx}$, which reverses its sign when the chemical potential is tuned from electron- to hole-doped regimes. By comparing the analytical results of the total $\chi_{yx}$ (black curve in Fig. 5c) with the numerical simulations (black circles in Fig. 5c), we find that the two-surface-state model gives a better description for the electron bands compared to the hole bands for Dirac surface states. This is because the Dirac surface states are usually close to the maximum of the bulk valence bands and thus the hole bands of surface states strongly hybridize with the bulk valence bands at finite momenta. This hybridization gives rise to the non-monotonic dispersion of the hole bands of surface states, leading to a complex momentum dependent contribution to $\chi_{yx}$ (Fig. 5a).

For $E_\text{F} \sim -0.15$ eV, the chemical potential starts to cross the first bulk valence quantum well subbands, and one more peak of $\chi_{yx}$ is seen, which shares the same sign of the surface state contribution in the hole doping regime. A strong hybridization between surface states and bulk quantum well states is expected in this regime. By further lowering the chemical potential to $E_\text{F} < -0.2$ eV, the second bulk valence quantum well subbands appear and their contribution to $\chi_{yx}$ has an opposite sign. By comparing our theory and experiments, we conclude that the major contribution to SOT magnetization switching observed in QAH insulators comes from both surface states and the first bulk valence quantum well subbands.

To summarize, we fabricate mesoscopic QAH sandwich Hall bar devices and realize electrical switching of the CEC chirality in QAH insulators by employing bulk and/or surface carriers induced SOT effect, which was theoretically simulated by a four-band model in a sandwich heterostructure configuration. Despite the low critical temperature of 5~10 K in modulation-doped QAH sandwiches [1] (Extended Data Figs. 1b and 6b), we can still implement the CEC electrical control protocol for topological quantum computation applications and proof-of-concept CEC-based electronic and spintronic devices. Moreover, our work provides new insights into the fundamental properties of the QAH insulators, e.g. the interplay between magnetism and topology, which may help tackle environmental challenges by revolutionizing the next generation of quantum material-based computation and information technology.

**Methods**

**MBE Growth of QAH sandwich heterostructures**

The 3-4-3 QAH sandwich heterostructures, specifically 3QL Cr-doped (Bi, Sb)$_2$Te$_3$ /4QL (Bi, Sb)$_2$Te$_3$/3QL Cr-doped (Bi, Sb)$_2$Te$_3$, are fabricated in a commercial MBE system (Omicron Lab10) with a base pressure lower than $2 \times 10^{-10}$ mbar. Before the growth of the QAH samples, the heat-treated insulating SrTiO$_3$ (111) substrates are first outgassed at 600 °C for 1 hour. Next, high-purity Bi(6N), Sb(6N), Cr(5N), and Te(6N) are evaporated from Knudsen effusion cells. During the growth of the QAH samples, the substrate is maintained at ~230 °C. The flux ratio of Te per (Bi + Sb + Cr) is set to greater than ~10 to prevent Te deficiency in the films. The growth rate for the QAH sandwiches is ~0.2 QL per minute. To avoid possible contamination and degradation in ambient circumstances, a ~3 nm thick Al$_2$O$_3$ layer is deposited on top of the QAH sandwich heterostructures in an atomic layer deposition system as soon as we take them out of the MBE chamber.

**Electron beam lithography of the QAH Hall bar devices**

The QAH sandwich heterostructures with a ~3 nm thick Al$_2$O$_3$ capping layer are patterned into Hall bar devices with the width *w* of 1~10 μm using two-step electron beam lithography. The aspect ratio of all QAH Hall bar devices used in this work is ~4. We use Cr (10 nm)/Au (30 nm) layers as the electrodes of these QAH Hall bar devices. Before the deposition of the Cr/Au electrodes using electron beam evaporation, the ~3 nm Al$_2$O$_3$ layer is etched using CD26. Finally, we use Ar plasma to etch the QAH sandwich into the Hall bar geometry. To avoid the charging effect, we deposit a ~20 nm Au layer on top of the resist before electron beam lithography.

**Electrical transport measurements**

Transport measurements are conducted using both a Physical Property Measurement System (Quantum Design DynaCool, 1.7 K, 9 T) and a dilution refrigerator (Bluefors, 10 mK, 9-1-1 T). The bottom gate voltage $V_g$ is applied using a Keithley 2450 meter. The excitation currents used in the PPMS (≥1.7 K) and dilution (~20 mK) measurements are 10 nA and 1 nA, respectively. The results reported here have been reproduced in ~10 Hall bar devices with a width from 1 to 10 μm in our PPMS and ~5 Hall bar devices with a width from 1 to 5 μm in our dilution refrigerator (Extended Data Fig. 10). The current pulse with a duration of 5 milliseconds, parallel to the in-plane magnetic field, is applied using a Keithley 6221 source meter. The current pulse density $J_{\text{pluse}}$ used for magnetization switching is similar in all devices of different widths. More transport results are found in Extended Data Figs. 1 to 10 and Supplementary Information.

**Numerical simulations on SOT in QAH sandwich heterostructures**

To numerically solve the spectrum of the four-band Hamiltonian, we use the basis

$$\psi(z) = \langle z | \mathbf{k}_\parallel N \lambda \rangle = \sqrt{2/L} \sin N\pi z/L \, |\mathbf{k}_\parallel \lambda\rangle.$$

$\lambda$ is the orbital and spin degrees of freedom. $\sqrt{2/L} \sin N\pi z/L$ with $N$ as a positive integer satisfies the open boundary condition $\psi(z = 0) = \psi(z = L) = 0$ in the $z$-direction of a thin film. For $L = 10$ nm in our QAH sandwiches, $N$ is truncated at $N_{max} = 30$ that is large enough to reproduce the gapless Dirac surface states for $M = 0$ eV.

From the Hamiltonian, the response function of spin polarization can be obtained from the Kubo formula [38]

$$\chi_{yx} = e\hbar \sum_n \int \frac{d^2 \mathbf{k}_\parallel}{(2\pi)^2} \chi_{yx}(\mathbf{k}_\parallel, n, E_F) =$$

$$= -\frac{e\hbar}{4\pi} \sum_n \int \frac{d^2\bm{k}_\parallel}{(2\pi)^2} \frac{2\Gamma^2}{\left(\left(E_F - E_{\bm{k}_\parallel n}\right)^2 + \Gamma^2\right)^2} \langle \bm{k}_\parallel n | \sigma_y | \bm{k}_\parallel n \rangle \langle \bm{k}_\parallel n | v_x | \bm{k}_\parallel n \rangle.$$

Here $E_{\bm{k}_\parallel n}$ and $|\bm{k}_\parallel n\rangle$ are the $n^{\text{th}}$ energies and eigenstates of Hamiltonian. The contribution on Fermi surfaces is defined as $\chi_{yx}(k_x n, E_F = E_{\bm{k}_\parallel n})$ with $\bm{k}_\parallel = k_x \hat{x}$ taken as the typical point on the Fermi surfaces. $\Gamma$ is the self-energy broadening from disorder. $\Gamma = 0.03$ eV are used. In Supplementary Information, we also perform the numerical simulations on the Landau-Lifshitz-Gilbert equation to show the effects of $\bm{\tau}_{\text{so,ISGE}}$ in magnetization switch.

To further understand $\chi_{yx}$, the two-surface-state model is used with the Hamiltonian

$$H_{2s} = \begin{pmatrix} -v_f(k_y \sigma_x - k_x \sigma_y) + M\sigma_z + V_0/2 & 0 \\ 0 & v_f(k_y \sigma_x - k_x \sigma_y) + M\sigma_z - V_0/2 \end{pmatrix}.$$

The first (second) diagonal block is the top (bottom) surface. $v_f$ is the Fermi velocity, $M$ is the magnetization and $V_0$ is the asymmetric potential between two surfaces. $v_f = 3.7$ eVÅ, $M = 0.041$ eV, and $V_0 = 0.032$ eV are fitted from the surface state spectrum of $E_F > -0.15$ eV in Fig. 5a.

**Acknowledgments:** We are grateful to Y. -T. Cui, I. Garate, L.-Q. Liu, N. Samarth, W. -D. Wu, D. Xiao, and X. -D. Xu for helpful discussions. This work is primarily supported by ARO Young Investigator Program Award (W911NF1810198), including sample synthesis and device fabrication. The PPMS measurements are supported by the AFOSR grant (FA9550-21-1-0177) and the NSF-CAREER award (DMR-1847811). The theoretical calculations and simulations are supported by the Penn State MRSEC for Nanoscale Science (DMR-2011839). C. Z. C. acknowledges the support from Gordon and Betty Moore Foundation's EPiQS Initiative (GBMF9063 to C. Z. C.).

**Author contributions:** C. -Z. C. conceived and designed the experiment. Y. -F. Z. grew the QAH sandwich samples. L.-J. Z fabricated the Hall bar devices using electron beam lithography. W. Y., L.-J. Z, R. Z., and M. K. performed electrical transport measurements. K. Y., R. M., and C.-X. L. did numerical simulations and provided theoretical support. W. Y., K. Y., C.-X. L., and C.-Z. C. analyzed the data and wrote the manuscript with inputs from all authors.

**Competing interests:** The authors declare no competing financial interests.

**Data availability:** The datasets generated during and/or analyzed during this study are available from the corresponding author upon reasonable request.

**Figures and figure captions:**

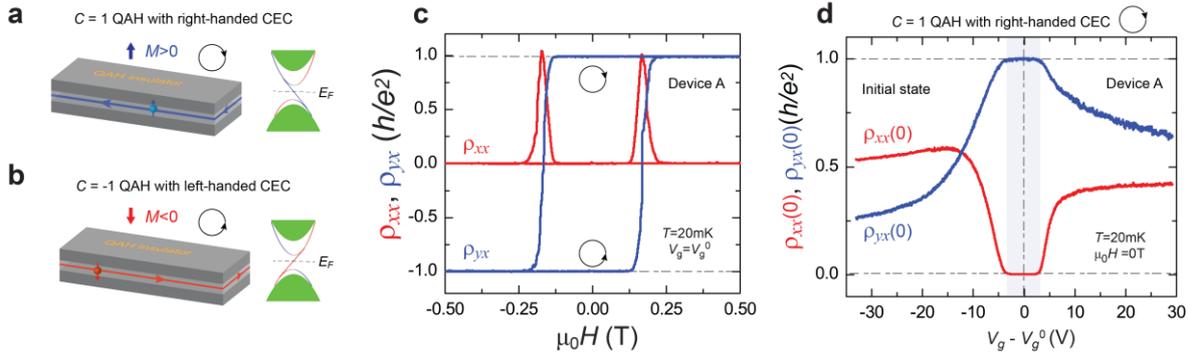

**Fig. 1| Edge current chirality in QAH insulators. a**, **b**, Schematics of right- and left-handed CEC in QAH sandwiches, specifically 3QL Cr-doped $(Bi,Sb)_2Te_3$/4QL $(Bi,Sb)_2Te_3$/3QL Cr-doped $(Bi,Sb)_2Te_3$. The CEC is shown in the real space (left) and momentum space (right) for positive and negative magnetization, respectively. **c**, Magnetic field $\mu_0H$ dependence of the longitudinal resistance $\rho_{xx}$ (red) and the Hall resistance $\rho_{yx}$ (blue) measured at $V_g = V_g^0 = +0.8$ V and $T = 20$ mK. $\rho_{yx}(0)$ can be used as a probe to identify CEC chirality. $\rho_{yx}(0) \sim +h/e^2$ and $\rho_{yx}(0) \sim -h/e^2$ correspond to right- and left-handed CEC, respectively. **d**, Gate ($V_g - V_g^0$) dependence of $\rho_{yx}(0)$ (blue) and $\rho_{xx}(0)$ (red) of the $C=1$ QAH insulator with right-handed CEC at zero magnetic field and $T = 20$ mK.

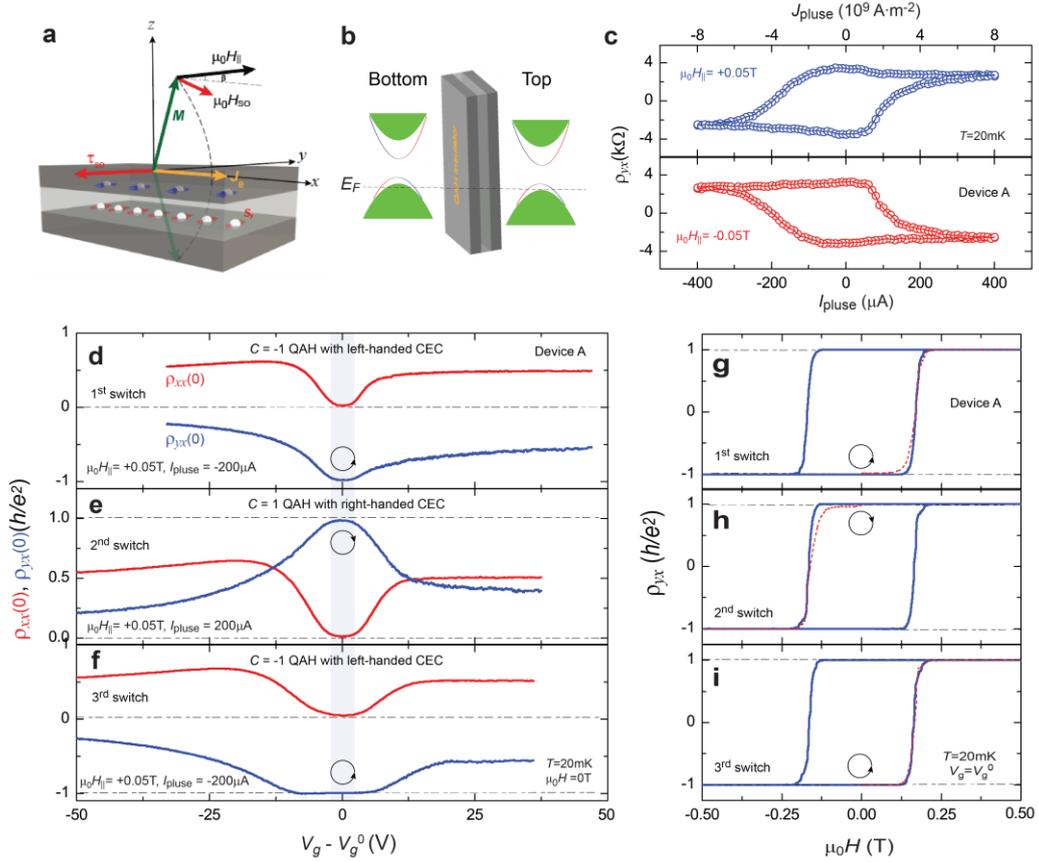

**Fig. 2| Electrical switching of edge current chirality in QAH insulators through bulk and surface carriers generated SOT. a**, Schematic of the current-pulse-induced magnetization switching in a QAH sandwich heterostructure under an in-plane external magnetic field. The angle between the *x*-axis and the direction of the in-plane external magnetic field |β| is less than 4°. **b,** Schematic of the magnetic TI sandwich heterostructure with symmetry-inequivalent top and bottom surfaces when applying a bottom gate voltage $V_g$. The chemical potential is tuned to cross both bulk valence bands and surface states. **c,** Current pulse $I_{pulse}$ dependence of the Hall resistance $\rho_{yx}$ under an in-plane magnetic field $\mu_0 H_\parallel = \pm 0.05$ T at $T = 20$ mK. The corresponding current pulse density $J_{pulse}$ is shown on the upper horizontal axis. **d-f,** Gate ($V_g - V_g^0$) dependence of $\rho_{yx}(0)$ (blue) and $\rho_{xx}(0)$ (red) of the QAH insulator after the first switch with $I_{pulse} \sim -200$ μA (**d**), second switch with $I_{pulse} \sim 200$ μA (**e**), and third switch with $I_{pulse} \sim -200$ μA (**f**) under $\mu_0 H_\parallel = +0.05$ T. **g-i,** $\mu_0 H$ dependence of the Hall resistance $\rho_{yx}$ at $V_g = V_g^0$ and $T = 20$ mK after the first (**g**), second (**h**), and third (**i**) switches. The red dashed curves correspond to the initial magnetization process after each switch.

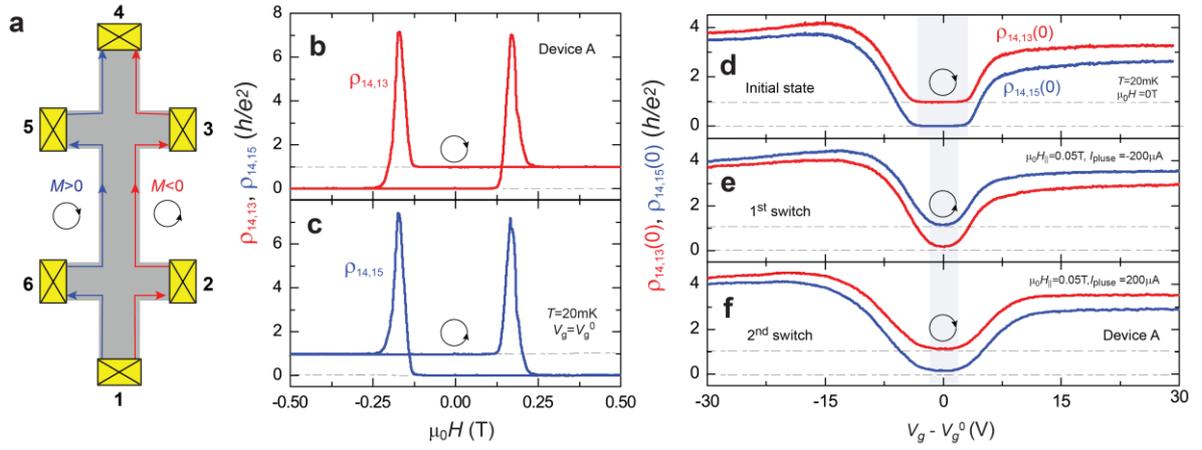

**Fig. 3| Demonstration of edge current chirality switching in QAH insulators through three-terminal measurements. a**, Schematic of chiral edge channels when the current flows from 1 to 4. The blue and red lines indicate right- and left-handed CEC for positive and negative magnetization, respectively. **b, c**, $\mu_0 H$ dependence of the three-terminal resistances $\rho_{14,13}$ (**b**) and $\rho_{14,15}$ (**c**) at $V_g = V_g^0$ and $T = 20$ mK. **d-f,** Gate ($V_g - V_g^0$) dependence of $\rho_{14,13}(0)$ (red) and $\rho_{14,15}(0)$ (blue) of the QAH insulator at the initial state (**d**), after the first (**e**) and second (**f**) switches.

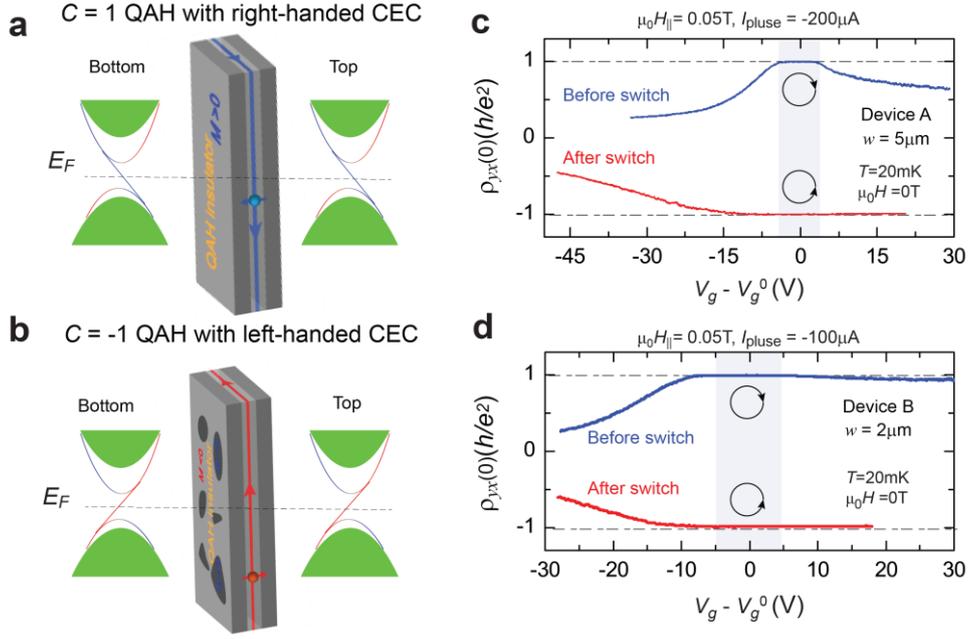

**Fig. 4| Electrical switching of edge current chirality in QAH insulators through surface carriers generated SOT. a**, Schematic of the initial $C = 1$ QAH sandwich heterostructure with fully-aligned positive magnetization. **b**, Schematic of the $C = -1$ QAH sandwich heterostructure with percolating negative magnetization after SOT switching at $V_g = V_g^0$. The dark grey areas depict magnetic domains that fail to switch. Due to the quantum percolation property, the sample still shows a well-quantized $C=-1$ QAH effect. **c**, Gate ($V_g - V_g^0$) dependence of $\rho_{yx}(0)$ of Device A ($w = 5$ μm) before (blue) and after (red) magnetization switching at $V_g = V_g^0$ with $I_{pulse} \sim -200$ μA under $\mu_0 H_\| = +0.05$ T at $T = 20$ mK. **d,** Same to (**c**) but for Device B ($w = 2$ um), $I_{pulse}$ used for magnetization switching is $\sim -100$ μA.

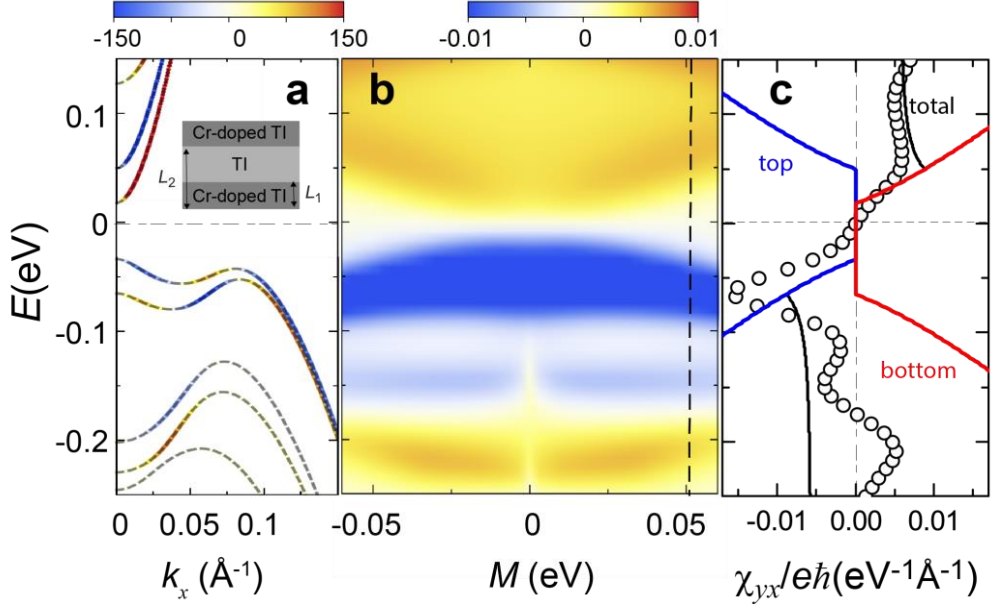

**Fig. 5| Theoretical calculations of $\chi_{yx}$ in the 3-4-3 QAH sandwich heterostructure. a,** Energy spectrum for the 3-4-3 QAH sandwich heterostructure with $M = 0.05$ eV. The color represents the contribution $\chi_{yx}(k_x, n, E_F = E_{k_x n})$ when the Fermi surface crosses the $n^{th}$ band at $k_x$. **b,** The spin response function $\chi_{yx}$ for the 3-4-3 QAH sandwich heterostructure under different magnetization $M$ and chemical potential $E_F$. The color represents the final spin response function $\chi_{yx}/e\hbar$. **c,** Comparison of $\chi_{yx}$ from the two-surface-state model and four-band Hamiltonian of the 3-4-3 QAH sandwich heterostructure. The blue (red) lines are $\chi_{yx}$ from the top (bottom) surface. The black line is the total of the two surfaces, while the black circles are calculated from the 3-4-3 QAH sandwich model with $M = 0.05$ eV, corresponding to the black vertical dashed line in (**b**). Note that the vertical axis in (**b**) and (**c**) is the chemical potential $E_F$.

**Extended data figures and figure captions:**

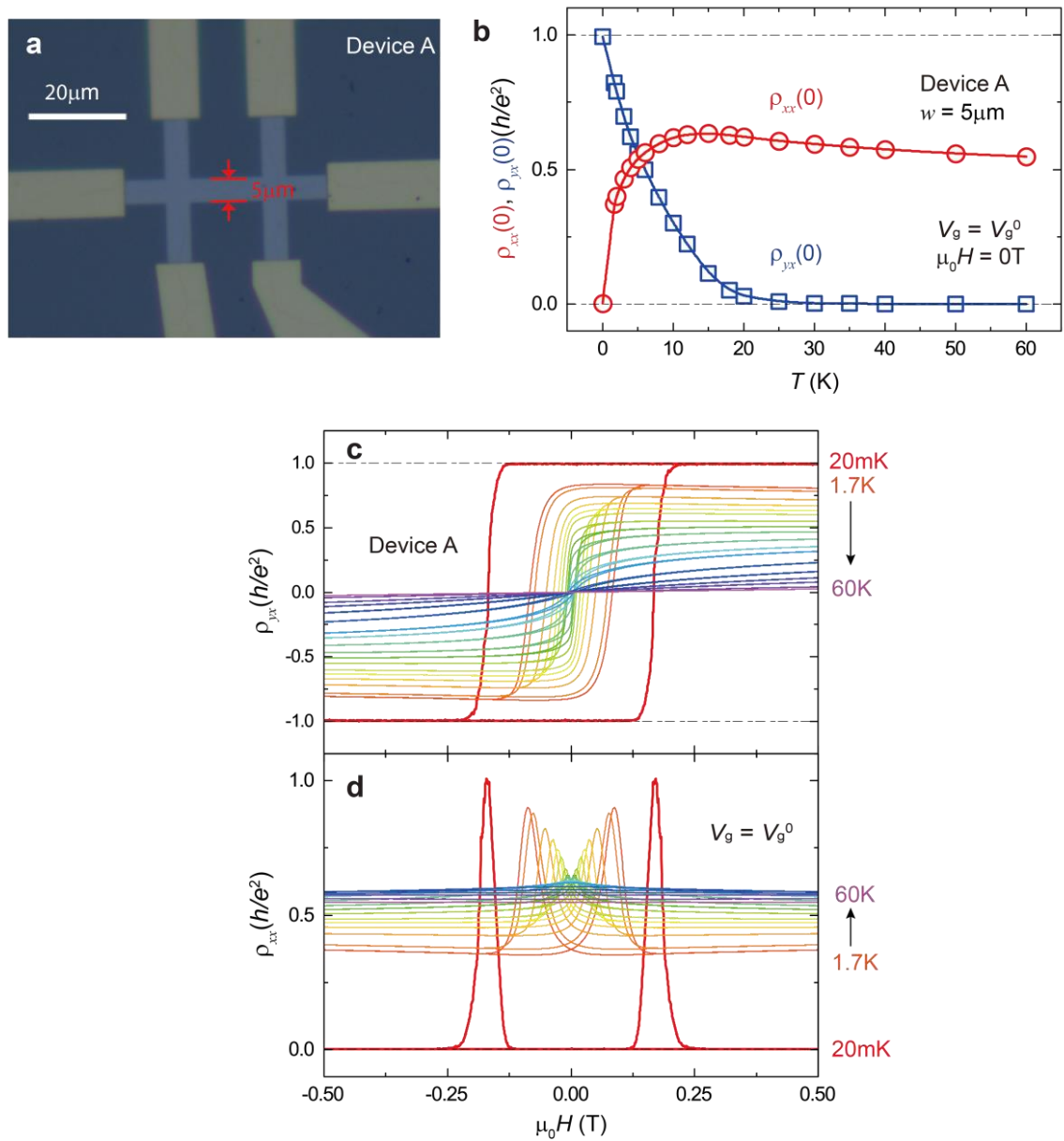

**Extended Data Fig. 1| QAH state in Device A ($w$ =5 μm). a,** The optical photograph of Device A. The effective area of the Hall bar device is 20 μm × 5 μm. **b,** Temperature dependence of $\rho_{yx}(0)$ (blue squares) and $\rho_{xx}(0)$ (red circles) at $V_g = V_g^0$. All measurements are taken at $\mu_0 H = $ 0T after magnetic field training. The critical temperature of the QAH state in Device A is ~5.3 K. **c, d,** $\mu_0 H$ dependence of $\rho_{yx}$ and $\rho_{xx}$ measured at different temperatures and $V_g = V_g^0$.

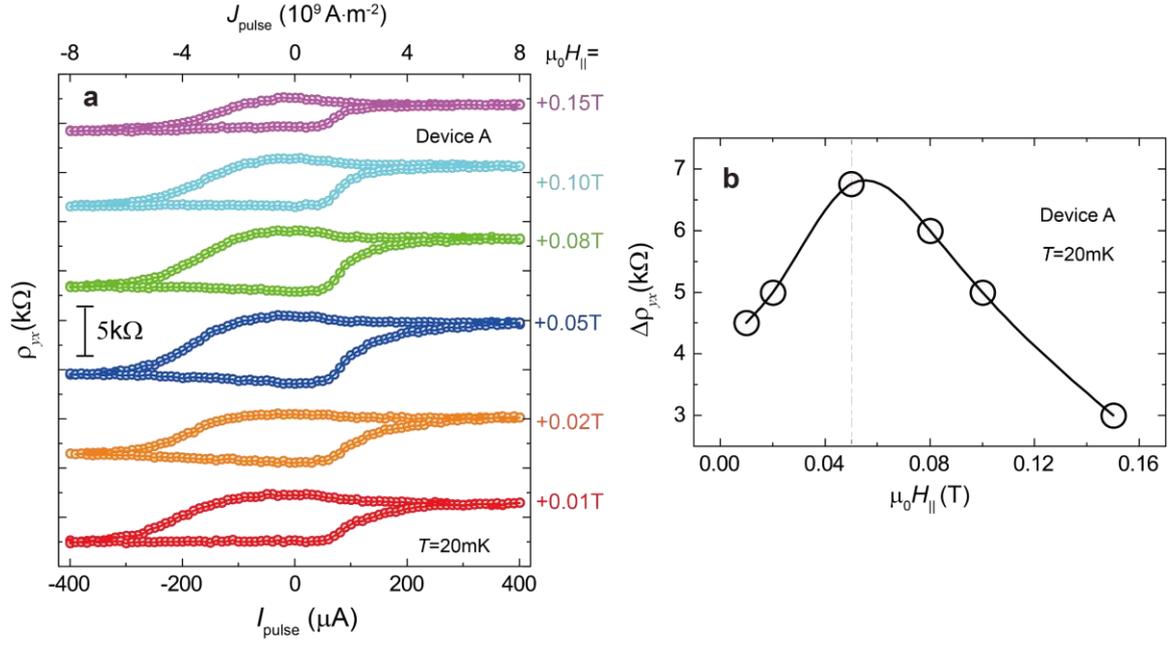

**Extended Data Fig. 2| SOT-induced magnetization switching under different in-plane magnetic fields in Device A. a**, Current pulse $I_{pulse}$ dependence of $\rho_{yx}$ under different $\mu_0 H_\parallel$ at $T$=20 mK. The corresponding current pulse density $J_{pulse}$ is shown on the upper horizontal axis. The hysteresis loops reflect the reversal of magnetization direction. All these SOT switching measurements are done at $\rho_{yx}$ ~0.27 $h/e^2$. The data curves are vertically shifted for clarity. **b**, $\mu_0 H_\parallel$ dependence of the Hall resistance change $\Delta\rho_{yx}$ at $T$=20 mK. $\Delta\rho_{yx}$ is maximized near $\mu_0 H_\parallel$ = +0.05 T, so we choose $|\mu_0 H_\parallel|$ =0.05 T for the SOT-induced magnetization switching in Device A.

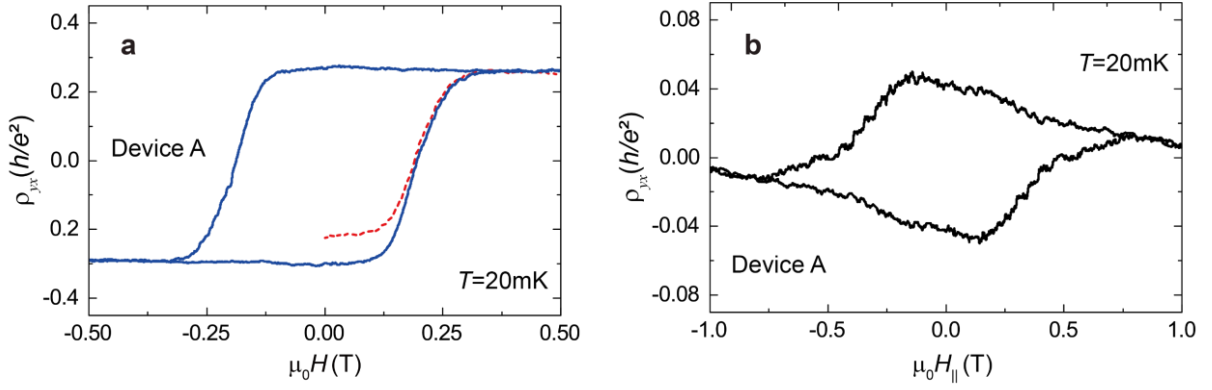

**Extended Data Fig. 3| More magneto-transport properties of Device A. a**, $\mu_0 H$ dependence of $\rho_{yx}$ without tuning $V_g$ after SOT switching at $\rho_{yx}(0)$ ~0.27 $h/e^2$ and $T$ =20 mK. After SOT-induced switching, $\rho_{yx}(0)$ is ~ -0.225 $h/e^2$. After applying $\mu_0 H$ ~0.5 T to align the magnetization, $\rho_{yx}(0)$ ~ 0.268 $h/e^2$. Therefore, the SOT magnetization switching ratio at $\rho_{yx}(0)$ ~0.27 $h/e^2$ by applying $|I_{\text{pulse}}|$ ~200 μA under $\mu_0 H_{//}$ = +0.05 T is ~83.8%. When the sample magnetization is fully aligned, the negligible $\rho_{yx}$ difference suggests the gating effect induced by the injection of $I_{\text{pulse}}$ is much weaker when the SOT switching is done at $|\rho_{yx}(0)|$ ~0.27 $h/e^2$. The red dashed curve corresponds to the initial magnetization process after SOT switching. **b**, $\mu_0 H_{||}$ dependence of $\rho_{yx}$ at $T$ =20 mK when the sample is tuned to $\rho_{yx}(0)$ ~0.27 $h/e^2$. We find that the anisotropy field $K$ is ~0.7 T and thus the sample magnetization almost points upward and downward under $|\mu_0 H_{//}|$ =0.05 T.

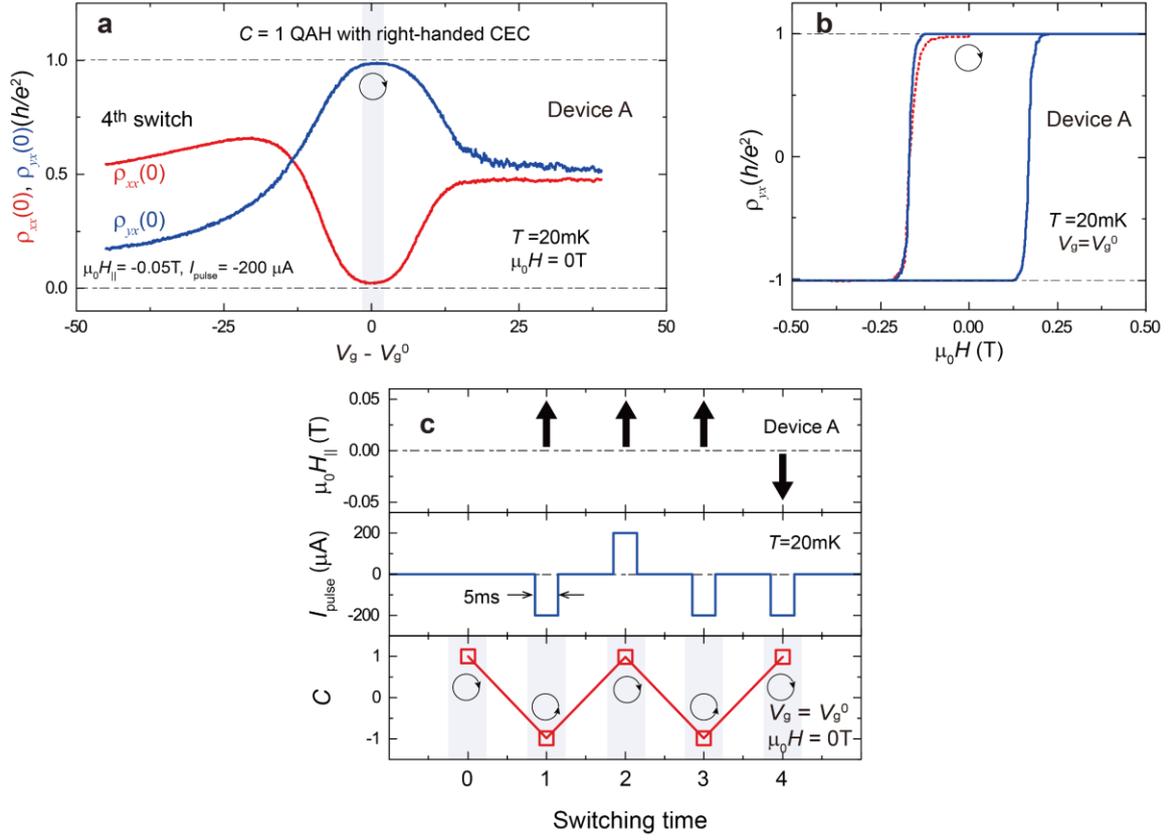

**Extended Data Fig. 4| Electrical switching of edge current chirality in Device A through bulk and surface carriers generated SOT. a,** Gate ($V_g$-$V_g^0$) dependence of $\rho_{yx}(0)$ (blue) and $\rho_{xx}(0)$ (red) of the QAH insulator after the fourth switch with $I_{pulse} \sim$ -200 μA under $\mu_0 H_{\parallel}$ = -0.05 T. The SOT switching is done at $\rho_{yx}(0) \sim$ -0.27 $h/e^2$ and $T$ =20 mK. **b,** $\mu_0 H$ dependence of $\rho_{yx}$ at $V_g = V_g^0$ and $T$ =20 mK after the fourth switch. The red dashed curve corresponds to the initial magnetization process after the SOT switching. **c,** Summary of all four switches of CEC chirality at $T$ =20 mK. The CEC chirality can be switched by changing the direction of either the in-plane magnetic field or the current pulse.

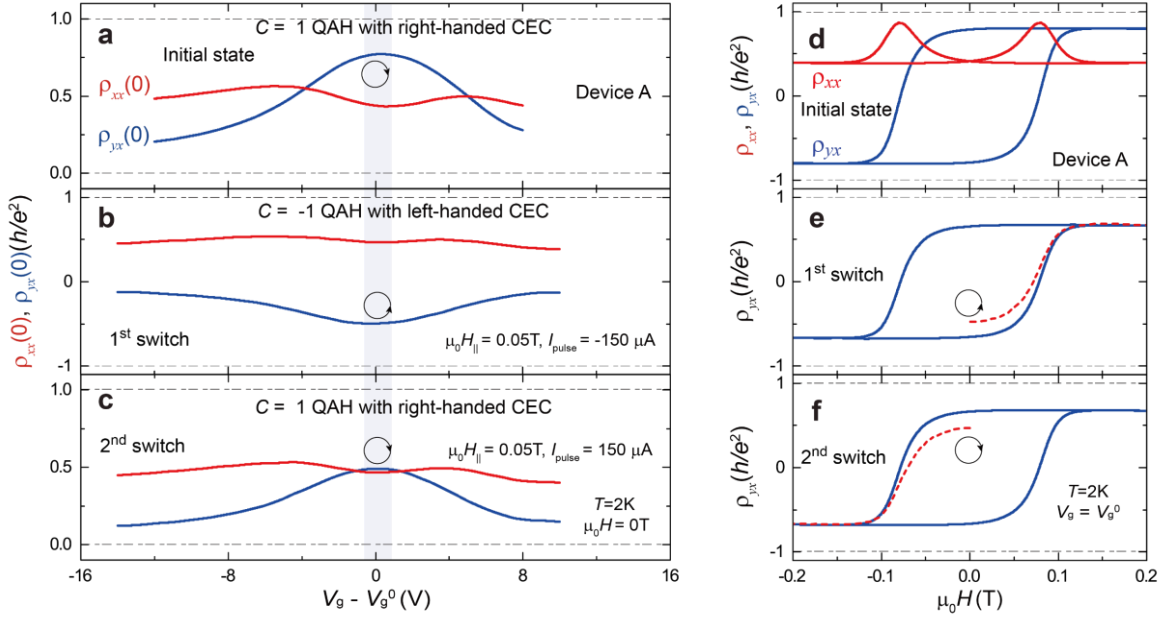

**Extended Data Fig. 5| Electrical switching of edge current chirality in Device A at $T$ =2 K. a-c,** Gate ($V_g$-$V_g^0$) dependence of $\rho_{yx}(0)$ (blue) and $\rho_{xx}(0)$ (red) of the QAH insulator with right-handed CEC (i.e. the initial state) (**a**), after the first (**b**) and second (**c**) switches. **d,** $\mu_0H$ dependence of $\rho_{yx}$ (blue) and $\rho_{xx}$ (red) at $V_g = V_g^0$ and $T$ =2 K. **e, f,** $\mu_0H$ dependence of $\rho_{yx}$ at $V_g = V_g^0$ and $T$ =2 K after the first (**e**) and second (**f**) switches, respectively. The red dashed curves correspond to the initial magnetization process after each switch.

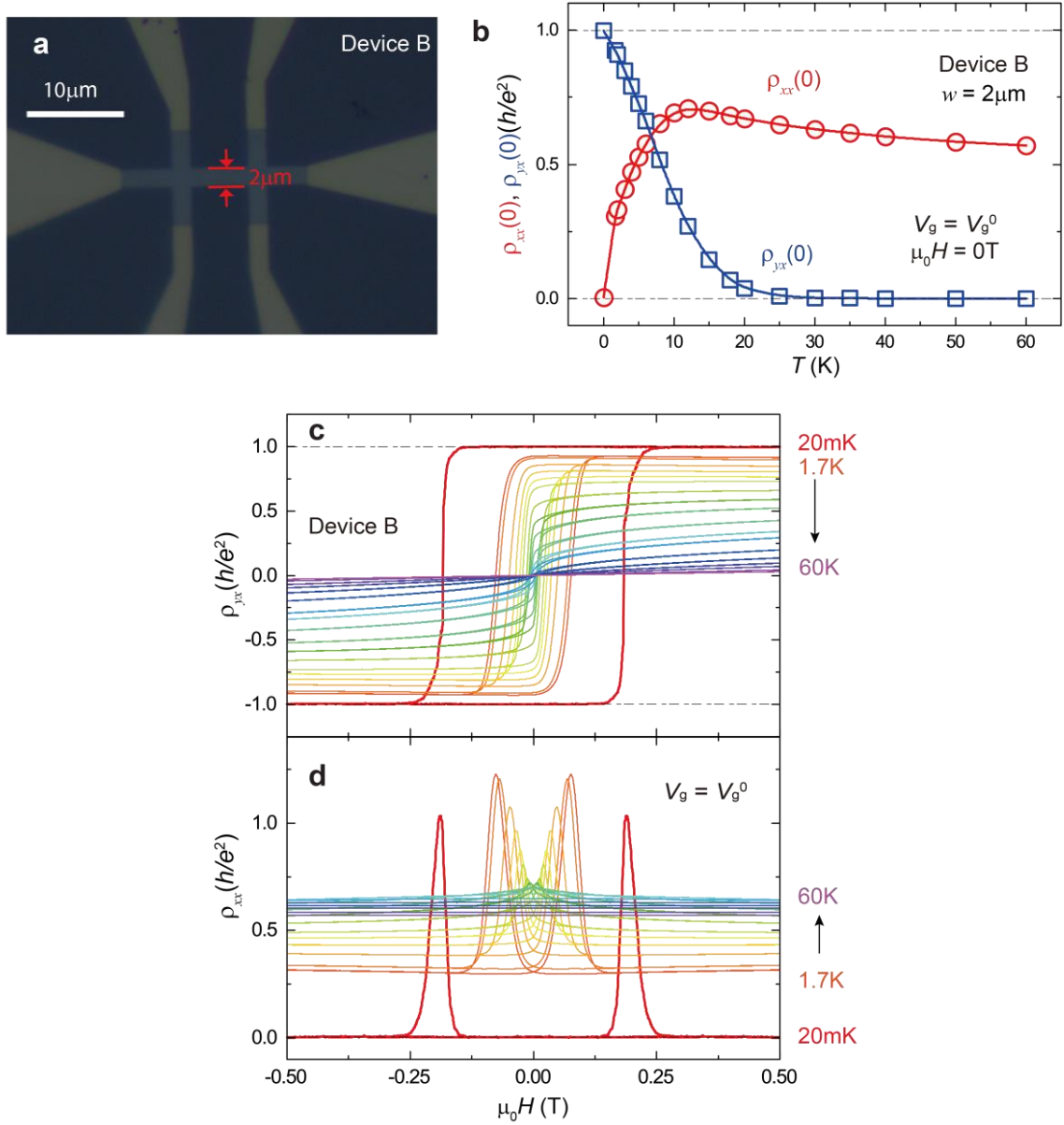

**Extended Data Fig. 6| QAH state in Device B ($w$ =2 μm). a,** The optical photograph of Device B. The effective area of the Hall bar device is 8 μm × 2 μm. **b,** Temperature dependence of $\rho_{yx}(0)$ (blue squares) and $\rho_{xx}(0)$ (red circles) at $V_g = V_g^0$. All measurements are taken at $\mu_0 H = 0$ T after magnetic field training. The critical temperature of the QAH state in Device A is ~6.8 K. **c, d,** $\mu_0 H$ dependence of $\rho_{yx}$ and $\rho_{xx}$ measured at different temperatures and $V_g = V_g^0$.

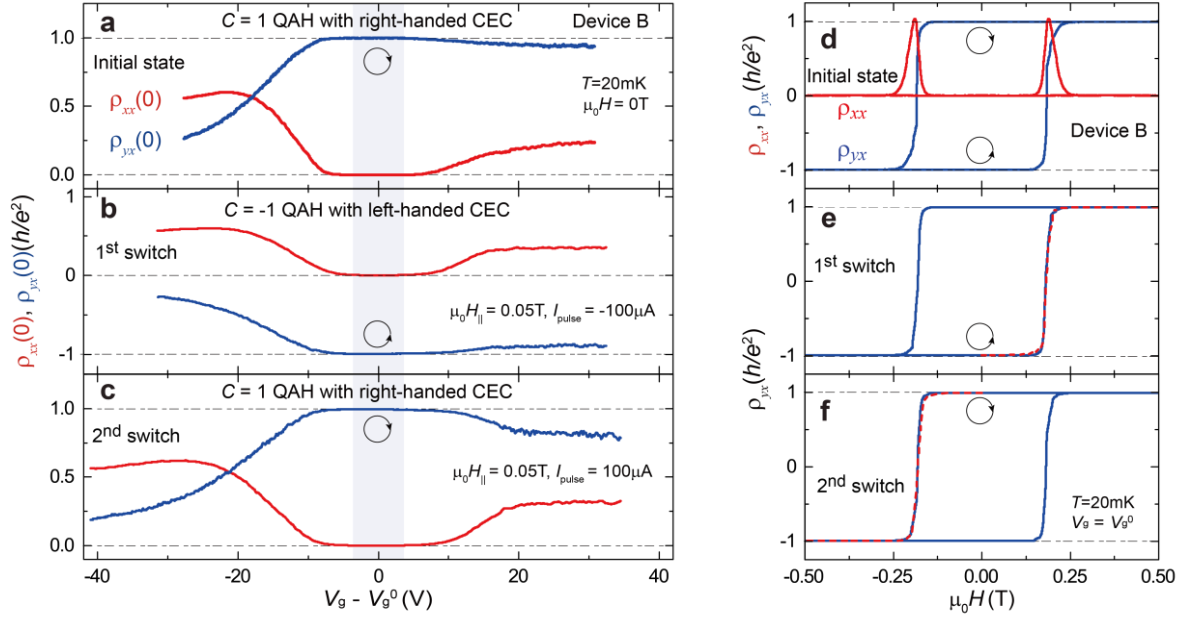

**Extended Data Fig. 7| Electrical switching of edge current chirality in Device B through bulk and surface carriers generated SOT. a-c,** Gate ($V_g$-$V_g^0$) dependence of $\rho_{yx}(0)$ (blue) and $\rho_{xx}(0)$ (red) of the QAH insulator with the right-handed CEC (i.e. the initial state) (**a**), after the first switch with $I_{pulse} \sim -100$ μA (**b**) and the second switch with $I_{pulse} \sim 100$ μA (**c**) under $\mu_0 H_\parallel \sim +0.05$ T. Both SOT switches are done at $|\rho_{yx}(0)| \sim 0.27$ $h/e^2$ and $T=20$ mK. **d,** $\mu_0 H$ dependence of $\rho_{yx}$ (blue) and $\rho_{xx}$ (red) at $V_g = V_g^0$ and $T=20$ mK. **e, f,** $\mu_0 H$ dependence of $\rho_{yx}$ at $V_g = V_g^0$ and $T=20$ mK after the first (**e**) and second (**f**) switches. The red dashed curves correspond to the initial magnetization process after each switch.

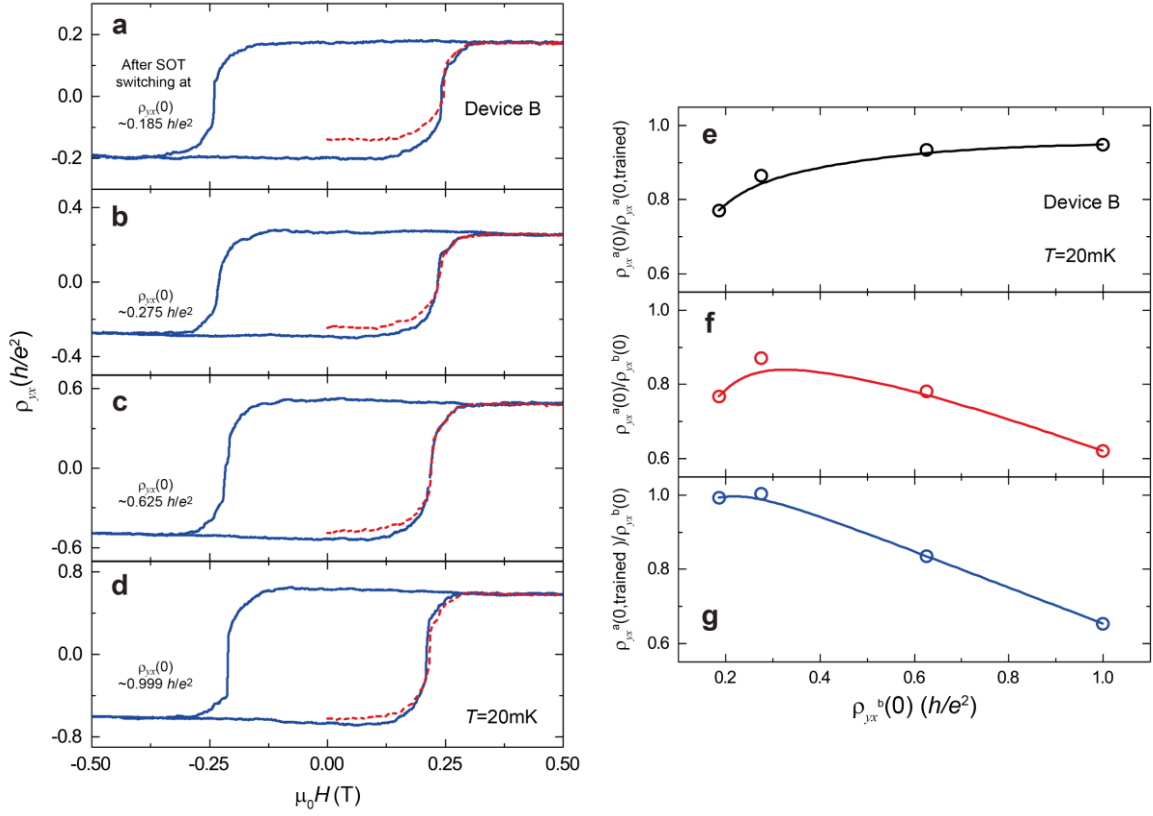

**Extended Data Fig. 8| More magneto-transport properties of Device B after SOT switching. a-d**, $\mu_0 H$ dependence of $\rho_{yx}$ without tuning $V_g$ after SOT switching at $\rho_{yx}(0) \sim 0.185$ $h/e^2$ (**a**), ~0.275 $h/e^2$ (**b**), ~0.625 $h/e^2$ (**c**), and ~0.999 $h/e^2$ (**d**), respectively. After SOT-induced switching, the corresponding $\rho_{yx}(0)$ is ~ -0.142 $h/e^2$ (**a**), ~ -0.240 $h/e^2$ (**b**), ~ -0.489 $h/e^2$ (**c**), and ~ -0.620 $h/e^2$ (**d**), respectively. After applying $\mu_0 H \sim 0.5$ T to align the magnetization, $\rho_{yx}(0)$ becomes ~0.184 $h/e^2$ (**a**), ~0.277 $h/e^2$ (**b**), ~0.523 $h/e^2$ (**c**), and ~0.653 $h/e^2$ (**d**), respectively. The red dashed curve corresponds to the initial magnetization process after each switch. **e-g,** Three ratios $\rho_{yx}^a(0)/\rho_{yx}^a(0, \text{trained})$ (**e**), $\rho_{yx}^a(0)/\rho_{yx}^b(0)$ **f**), $\rho_{yx}^a(0, \text{trained})/\rho_{yx}^b(0)$ (**g**) as a function of $\rho_{yx}^b(0)$, where the SOT switching is done. $\rho_{yx}^b(0)$: the zero magnetic field Hall resistance before SOT switching. $\rho_{yx}^a(0)$: the zero magnetic field Hall resistance after SOT switching. $\rho_{yx}^a(0, \text{trained})$: the zero magnetic field Hall resistance after SOT switching and $\mu_0 H \sim 0.5$ T training. All measurements are taken at $T = 20$ mK. For the SOT switching done near the QAH regime, the $\rho_{yx}^a(0)/\rho_{yx}^b(0)$ ratio cannot be used to estimate the magnetization switching ratio since $\rho_{yx} \propto M$ becomes invalid.

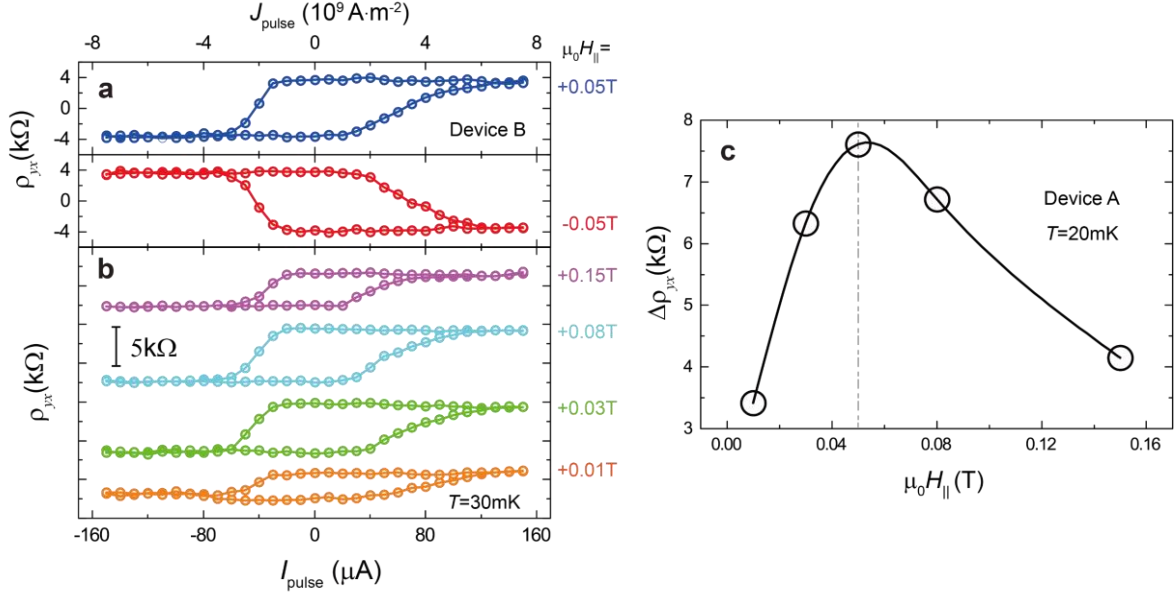

**Extended Data Fig. 9| SOT-induced magnetization switching under different in-plane magnetic fields in Device B ($w$ =2 μm). a,** Current pulse $I_{pulse}$ dependence of $\rho_{yx}$ under $\mu_0H_\parallel$ = +0.05 T(top) and $\mu_0H_\parallel$ = -0.05 T(bottom) at $T$=20 mK. **b,** $I_{pulse}$ dependence of $\rho_{yx}$ under different $\mu_0H_\parallel$ at $T$ =20 mK. The corresponding current pulse density $J_{pulse}$ in (**a**) and (**b**) is shown on the upper horizontal axis. All these SOT switching measurements in (**a**) and (**b**) are done at $\rho_{yx}$ ~0.27 $h/e^2$. The data curves in (b) are vertically shifted for clarity. **c,** $\mu_0H_\parallel$ dependence of the Hall resistance change $\Delta\rho_{yx}$ at $T$ =20 mK. $\Delta\rho_{yx}$ is maximized near $\mu_0H_\parallel$ = +0.05 T, so we chose $|\mu_0H_\parallel|$ =0.05 T for the SOT-induced magnetization switching in Device B. We find that the optimal $\mu_0H_\parallel$ for the SOT-induced magnetization switching is independent of the width of the QAH Hall bar device.

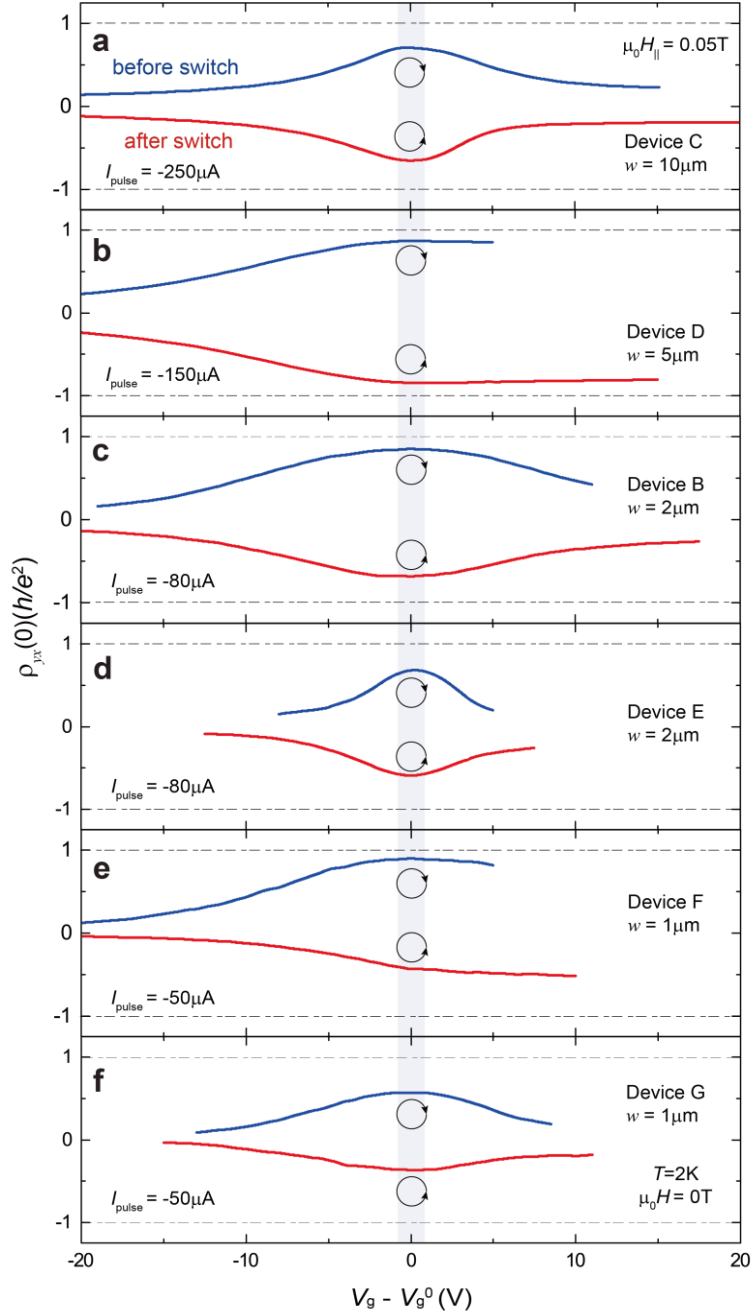

**Extended Data Fig. 10| Electrical switching of edge current chirality in more QAH insulator devices at $T$ =2 K. a-f,** Gate ($V_g$ - $V_g^0$) dependence of $\rho_{yx}(0)$ before (blue) and after (red) SOT magnetization switching in Device C ($w$ =10 μm) (**a**), Device D ($w$ =5 μm) (**b**), Device B ($w$ =2 μm) (**c**), Device E ($w$ =2 μm) (**d**), Device F ($w$ =1 μm) (**e**), and Device G ($w$ =1 μm) (**f**), respectively. The SOT-induced magnetization switching is all done at $\rho_{yx}(0)$~0.155 $h/e^2$ and $T$ =2 K under $\mu_0 H_\parallel$ ~ +0.05 T. The current pulse $I_{pulse}$ used for SOT magnetization switching becomes smaller with reducing the width of the QAH Hall bar device.